## Title

Spontaneous Zonal Symmetry Breaking of Tropical Rain Belt

## Authors

Tomoro Yanase[1,2,3], yanase@gsis.u-hyogo.ac.jp
Cathy Hohenegger[1], cathy.hohenegger@mpimet.mpg.de

## Affiliations

1. Max Planck Institute for Meteorology, Hamburg, Germany
2. Graduate School of Information Science, University of Hyogo, Kobe, Japan
3. RIKEN Center for Computational Science, Kobe, Japan

## Corresponding author

Correspondence to: Tomoro Yanase, yanase@gsis.u-hyogo.ac.jp

## Abstract

The intertropical convergence zone (ITCZ) is a central component of tropical climate, yet the conditions under which a tropical rain belt remains zonally extended or becomes unstable to zonal organization remain poorly understood. Here we investigate this problem using idealized nonrotating kilometer-scale simulations forced by a prescribed sea surface temperature (SST) distribution that varies only in the meridional direction. This setup produces an ITCZ-like rain belt while allowing spontaneous zonal convective self-aggregation (ZCSA) to emerge. A parameter sweep shows that ZCSA occurs preferentially when both the peak SST and the meridional SST amplitude are large. All ZCSA cases exhibit a temporary weakening of the meridional near-surface convergence that maintains the zonally elongated rain belt. Boundary-layer momentum and thermodynamic analyses show that this weakening is associated with enhanced lower-tropospheric stability over the cool subsiding region, which leads to a shallower boundary layer and stronger effective frictional damping of the meridional inflow. However, weak meridional convergence alone is not sufficient for ZCSA. Cases that develop ZCSA are additionally characterized by a large meridional contrast in moist static energy forcing, implying a strong demand for meridional energy transport. Consistent with this interpretation, ZCSA is accompanied by a reorganization of meridional moist static energy transport, including enhanced stationary eddy export from the warm region, and by a transition in which zonal moisture variability grows while the meridional moisture contrast weakens. These results suggest that zonal symmetry breaking of an ITCZ-like rain belt is favored when a weakened meridional inflow coincides with a large imposed meridional MSE-forcing contrast, and more generally highlight the coupled roles of convective thermodynamics, weak-temperature-gradient coupling, boundary-layer momentum balance, and large-scale energetic constraints in tropical rain-belt organization.

## 1. Introduction

The intertropical convergence zone (ITCZ), or more broadly the tropical rain belt, is a fundamental component of the tropical atmosphere. Its position, width, and strength shape the global distribution of precipitation and are closely tied to the Hadley circulation and the large-scale transport of energy and moisture (Donohoe et al. 2013; Lau and Kim 2015; Adam et al. 2016a,b; Byrne et al. 2018). At the same time, the ITCZ is not simply a zonally uniform band of rainfall. It exhibits substantial internal variability and rich dynamical structure, including pronounced zonal inhomogeneity in deep convection and surface convergence (Maloney and Shaman 2008; Klocke et al. 2017; Windmiller 2024; Windmiller and Stevens 2024). Growing evidence suggests that this internal organization is dynamically important: stronger zonal convective clustering is associated with systematic changes in rain belt structure, including a broader meridional precipitation distribution and, in some circumstances, a more pronounced double-peaked character (Popp and Bony 2019; Popp et al. 2020a,b). Understanding tropical rain belts therefore requires understanding not only their zonal-mean structure, but also how convection organizes within them.

Much of the current physical understanding of convective organization comes from studies of self-aggregation in radiative-convective equilibrium (RCE) under horizontally uniform boundary conditions. In that framework, idealized simulations have shown how interactions among convection, moisture, radiation, and circulation can generate large-scale spatial organization even in the absence of imposed horizontal heterogeneity (Nakajima and Matsuno 1988; Held et al. 1993; Tompkins and Craig 1998; Bretherton et al. 2005; Wing et al. 2017; Muller et al. 2022; Yanase et al. 2020, 2022a,b). However, the connection between self-aggregation in uniform RCE and the organization of convection in the real tropics remains unsettled (Holloway et al. 2017; Jakob et al. 2019; Hohenegger and Jakob 2020; Huang and Wu 2022; Tobin et al. 2012; Masunaga et al. 2021; Bony et al. 2020; Beucler et al. 2020). Outside the RCE framework, tropical convection evolves in the presence of sea surface temperature (SST) gradients, large-scale overturning circulations, and often rotation, all of which can influence where convection forms and how it organizes. For this reason, idealized moist-convection studies have increasingly moved toward more structured configurations, including prescribed-SST experiments designed to examine organized tropical circulations such as mock-Walker circulations (Grabowski et al. 2000; Bretherton et al. 2006; Sobel and Neelin 2006; Lutsko and Cronin 2024; Wing et al. 2024).

Within this broader hierarchy of idealized tropical experiments, Müller and Hohenegger (2020, hereafter MH20) provide a particularly relevant precedent for the present study. Using a nonrotating small aqua-planet with zonally uniform but meridionally varying SST, they found that the imposed meridional SST distribution produces an equatorial convergence line resembling an ITCZ-like rain belt, which can later break up and contract zonally as convection self-aggregates. In their experiments, the equatorial peak SST was held fixed while the meridional SST decay away from the equator was varied, so that the meridional SST gradient, the mean SST, and the minimum SST changed together. Although the transient evolution differed across cases, all of their simulations eventually reached a zonally aggregated regime, and larger SST contrasts tended to delay the onset of zonal aggregation and maintain longer convergence lines. These results clearly suggest that meridional SST forcing influences the evolution and zonal stability of an ITCZ-like rain belt. At the same time, they leave several important questions open. Because the meridional SST gradient and the mean thermal state vary simultaneously, the distinct roles of these factors remain difficult to isolate. More fundamentally, it remains unclear whether rain belts under other SST distributions would also always evolve toward a zonally aggregated regime, or whether distinct zonally extended and zonally aggregated regimes can coexist across a broader parameter space. Resolving this issue requires separating the different physical effects of the imposed SST distribution. The meridional SST gradient is expected to affect the rain belt through its influence on the meridional pressure gradient and the associated low-level circulation, whereas the peak SST is expected to affect atmospheric moisture content, and hence static stability through moist-adiabatic control. Because these controls act on different parts of the coupled circulation–moisture system, the preferred rain-belt structure cannot be inferred from the previous experiments alone.

One possible route toward answering this question comes from boundary-layer (BL) dynamics and low-level convergence. SST gradients are known to influence tropical low-level winds and convergence through pressure-gradient, frictional, and momentum-balance constraints (Lindzen and Nigam 1987; Stevens et al. 2002; Back and Bretherton 2009a,b; Praturi and Stevens 2026; Winkler et al. 2025). Since an ITCZ-like rain belt is maintained by low-level inflow from cooler, subsiding regions into warmer convective regions, the sensitivity of that inflow to the imposed SST distribution may be central to whether the rain belt remains zonally extended or becomes unstable to zonal aggregation. In this view, the imposed SST pattern affects organization not only by setting the background thermodynamic contrast, but also by shaping the circulation that sustains the rain belt.

A complementary perspective is provided by the energy and moisture transports associated with organized convection. Tropical rain belts play a central role in the meridional redistribution of moist static energy (MSE) and water vapor, and previous work has shown that stronger zonal convective clustering can be associated with weaker equatorial precipitation, reduced meridional moisture convergence into the equatorial zone, and a wider meridional rain distribution (Popp and Bony 2019). In aquaplanet simulations, different aspects of clustering have also been shown to influence different aspects of the tropical climate: reductions in the total convective area primarily affect the meridional distribution of humidity and precipitation, whereas changes in the number of convective regions more strongly influence zonal variance (Popp et al. 2020a). These results suggest that zonal aggregation should be viewed not only as a change in cloud geometry, but also as a reorganization of the large-scale transport of moisture and energy by the tropical circulation.

The problem may also be linked to recent conceptual work on interactions between moisture modes and the Hadley circulation. Adames and Mayta (2024) proposed that synoptic-scale moisture modes can extract energy from the zonal-mean meridional moisture gradient, weaken that gradient through eddy moisture fluxes, and thereby modulate the Hadley circulation. Relatedly, recent cloud-resolving work on Western Pacific ITCZ breakdown events showed that episodes of ITCZ breakdown are associated with anomalous moisture advection linked to equatorial Rossby-wave activity, together with more spatially homogeneous convection and weakened meridional moisture gradients (Casallas et al. 2026). From this perspective, zonal aggregation in an ITCZ-like rain belt may be interpreted as the emergence of east-west moisture inhomogeneity within a system whose mean state is sustained by meridional gradients of moisture and energy. This raises the possibility that spontaneous zonal organization and meridional overturning are not independent aspects of tropical variability, but coupled expressions of the same underlying moist dynamics. This perspective is also consistent with recent discussion of links between moisture-mode dynamics and self-aggregation in idealized settings (Yanase et al. 2025).

Motivated by these questions, this study investigates how spontaneous zonal organization emerges within a meridionally forced ITCZ-like rain belt. We ask how the onset of zonal aggregation depends on the imposed thermodynamic contrast, how it is mediated by the low-level circulation and convergence that sustain the rain belt, and how it feeds back onto the large-scale transport of MSE. By addressing these questions in an idealized framework, we seek to clarify when a tropical rain belt becomes unstable to zonal reorganization and how such reorganization fits into a broader picture of tropical variability spanning convective organization, overturning circulation, and moisture transport.

## 2. Methods

### 2.1 Numerical model and experiment design

We use a nonhydrostatic model SCALE-RM (Nishizawa et al. 2015; Sato et al. 2015), following the nonrotating idealized RCE framework (Yanase et al. 2020, 2022a,b). As in those previous studies, planetary rotation is neglected, and the physical parameterizations are kept unchanged, including radiation (Sekiguchi and Nakajima 2008), cloud microphysics (Tomita 2008), surface turbulent fluxes (Beljaars and Holtslag 1991), and subgrid-scale turbulence (Scotti et al. 1993; Brown et al. 1994). This choice allows us to isolate the effects of the imposed SST distribution on convective organization while maintaining consistency with our earlier idealized simulations.

The main difference from Yanase et al. (2020, 2022a,b) is that here the lower boundary condition includes a prescribed non-uniform meridional SST distribution. The basic idea is closely related to that of MH20, who examined zonal self-aggregation under meridionally varying SST. Here we build on that framework by exploring a broader SST parameter space more systematically. In particular, we impose an SST profile that varies sinusoidally only in the meridional direction,

$$\mathrm{SST}(y) = \mathrm{SST}_{\mathrm{ave}} - \mathrm{SST}_{\mathrm{amp}} \cos\left(\frac{2\pi y}{L_y}\right),$$

where $\mathrm{SST}_{\mathrm{ave}}$ is the domain-mean SST, $\mathrm{SST}_{\mathrm{amp}}$ is the imposed meridional SST amplitude, and $L_y$ is the meridional domain width. Throughout this paper, the terms zonal and *x*-direction, as well as meridional and *y*-direction, are used interchangeably. Because $L_y$ is fixed in this study, specifying $\mathrm{SST}_{\mathrm{ave}}$ and $\mathrm{SST}_{\mathrm{amp}}$ uniquely determines both the peak SST, $\mathrm{SST}_{\mathrm{max}} = \mathrm{SST}_{\mathrm{ave}} + \mathrm{SST}_{\mathrm{amp}}$, and the imposed meridional SST gradient. In the discussion below, we therefore organize the results primarily in terms of $\mathrm{SST}_{\mathrm{max}}$ and $\mathrm{SST}_{\mathrm{amp}}$.

We examine 12 runs in total. The peak SST, $\mathrm{SST}_{\mathrm{max}}$, is varied from 299 to 305 K at 2-K intervals, yielding four values, while $\mathrm{SST}_{\mathrm{amp}}$ is varied from 1 to 5 K at 2-K intervals, yielding three values. The combination of these parameters defines a two-dimensional parameter space spanning both the overall thermal state of the convecting region and the strength of the imposed meridional SST contrast. For brevity, each simulation case is referred to by a format "max$X$amp$Y$," where $X$ and $Y$ denote $\mathrm{SST}_{\mathrm{max}}$ (in K) and $\mathrm{SST}_{\mathrm{amp}}$ (in K), respectively; for example, the case with $\mathrm{SST}_{\mathrm{max}}$ = 301 K and $\mathrm{SST}_{\mathrm{amp}}$ = 1 K is denoted "max301amp1."

Each run is integrated for 200 days. The model output is available at hourly intervals. In the analyses below, daily averaging and low-pass filtering are applied where appropriate; the exact treatment is stated where relevant. Due to limitations in data storage size, the full 3D output is limited to the initial 50 days to observe the time evolution in the early phase and the final 30 days to observe the statistical equilibrium phase, and these are only used in the analysis related to BL momentum budget and MSE transport in Figs. 5 and 8.

The horizontal domain size is $L_x = 6144$ km and $L_y = 3072$ km, in the *x*-direction and *y*-direction, respectively, with a horizontal grid spacing of 4 km. A systematic assessment of sensitivity to domain size and horizontal resolution is clearly important, but it is not the focus of the present study. Nevertheless, the chosen domain is sufficiently large to accommodate mesoscale to large-scale convective organization: it exceeds the domain sizes used in several representative self-aggregation studies, such as $(768\ \mathrm{km})^2$ in Wing and Emanuel (2014), $(500\ \mathrm{km})^2$ in Tompkins and Semie (2017), and $(2823\ \mathrm{km})^2$ in MH20, and it is also larger than the characteristic length scale of about 3000 km suggested in Patrizio and Randall (2019) and Yanase et al. (2022b). The 4-km grid spacing is likewise of the same order as that used in related studies focusing on large-scale convective organization, for example the 3-km simulations of Patrizio and Randall (2019) and MH20.

## 3. Results

### 3.1 Regimes and temporal evolution of zonal self-aggregation

We begin by contrasting two representative simulations that exhibit clearly different horizontal structures of the tropical rain belt (Fig. 1). In the case with relatively low peak SST and weak meridional SST contrast (max301amp1), both outgoing longwave radiation (OLR) and precipitable water (PW) retains an approximately zonally extended structure. By contrast, in the case with high peak SST and strong meridional SST contrast (max305amp5), both OLR and PW exhibit pronounced zonal inhomogeneity, with convection and moisture concentrated in a localized zonal sector, which is similar to the horizontal distribution reported by MH20. We refer to this latter regime as zonal convective self-aggregation (ZCSA), and to the former regime as a non-ZCSA case. This contrast suggests that the ITCZ-like rain belt does not respond uniformly across the parameter space, but instead exhibits distinct structural regimes depending on the imposed SST conditions.

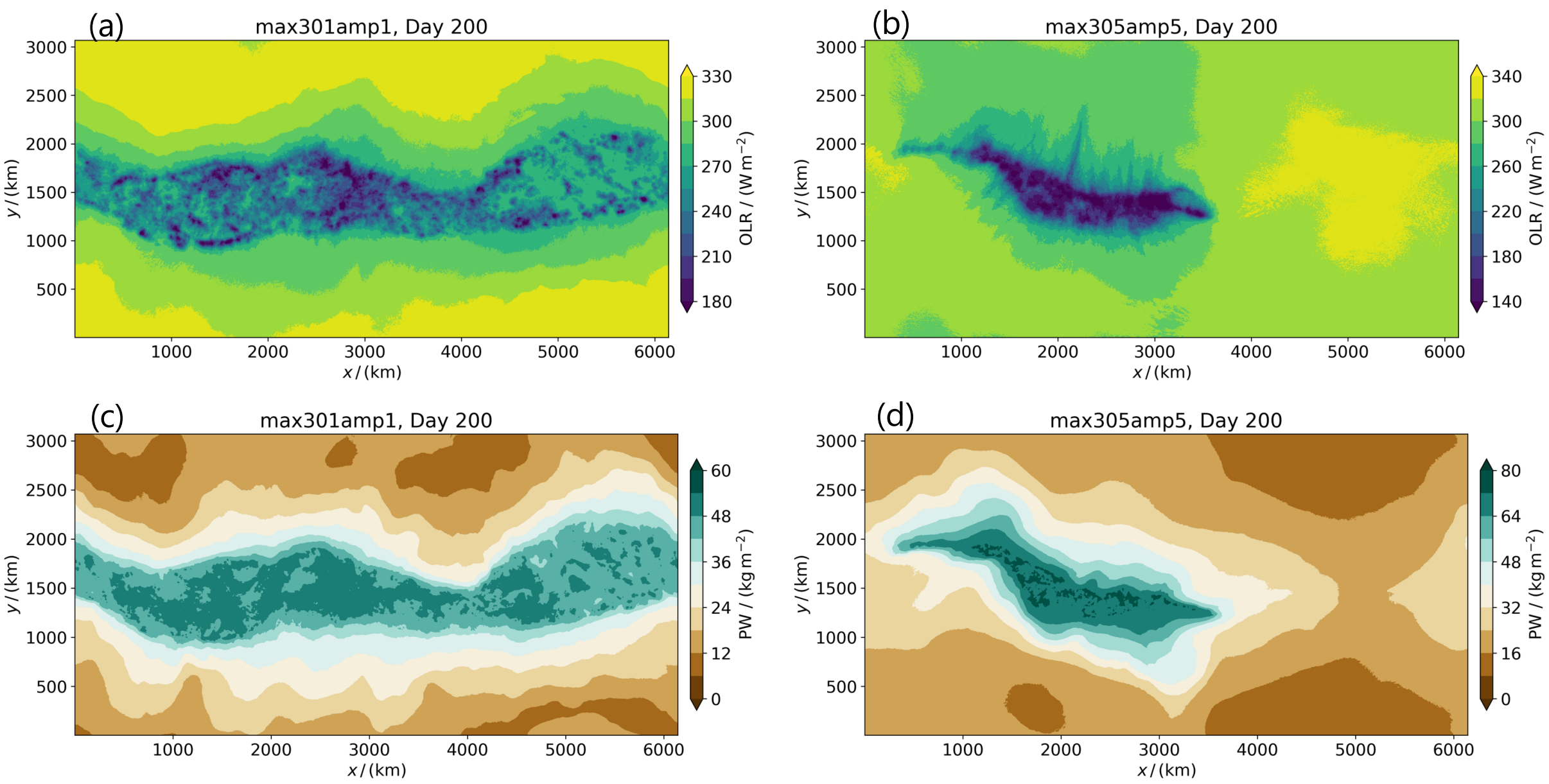

- Figure 1. Horizontal distribution of daily-mean OLR (a and b) and PW (c and d) on the last day of simulation. The panels on the (a) and (c) are for a case with $\mathrm{SST_{max}}$ of 301 K and $\mathrm{SST_{amp}}$ of 1 K (max301amp1). The panels on the (b) and (d) are for a case with $\mathrm{SST_{max}}$ of 305 K and $\mathrm{SST_{amp}}$ of 5 K (max305amp5).

To characterize the horizontal structure of the moisture field in a compact way, we introduce an anisotropy index based on PW, $\mathrm{AI_{PW}}$. It is defined as

$$\mathrm{AI_{PW}} = \frac{\mathrm{VarX_{PW}} - \mathrm{VarY_{PW}}}{\mathrm{VarX_{PW}} + \mathrm{VarY_{PW}}},$$

where $\mathrm{VarX_{PW}}$ is the meridional mean of the zonal variance of PW and $\mathrm{VarY_{PW}}$ is the zonal mean of the meridional variance of PW. The index ranges from $-1$ to 1: significantly negative values correspond to PW variability that is primarily meridional, as in a zonally extended rainbelt, while values near zero indicate that zonal variability has become comparable in magnitude to the meridional variability. Significantly positive values would represent cases in which zonal variability dominates over meridional variability, but no such cases are found in the present simulations.

The time series of $\mathrm{AI_{PW}}$ (Fig. 2a) show that all simulations start from a similar structural state. At early times (days 1–30), $\mathrm{AI_{PW}}$ approaches close to $-1$ for all runs, indicating that the developing rainbelt is initially close to zonally uniform and primarily organized in the meridional direction. The subsequent evolution then separates into distinct regimes. On the one hand, $\mathrm{AI_{PW}}$ remains substantially negative throughout the mature stage of the simulation, indicating persistence of a zonally elongated rainbelt. On the other hand, $\mathrm{AI_{PW}}$ rises toward near-zero values, reflecting the development of pronounced zonal moisture inhomogeneity. The contrast between these two classes is consistent with the representative horizontal patterns shown in Fig. 1. We use a value of $-1/3$ to distinguish the two groups, and regard cases with $-1/3 \leq \mathrm{AI_{PW}} < 1/3$ as ZCSA regime and those with $-1 \leq \mathrm{AI_{PW}} < -1/3$ as non-ZCSA regime.

Figure 2b summarizes this behavior over the $\mathrm{SST_{max}}$–$\mathrm{SST_{amp}}$ parameter space using the mean $\mathrm{AI_{PW}}$ over the final 30 days of each 200-day simulation. The distribution clearly shows that ZCSA is favored when both $\mathrm{SST_{max}}$ and $\mathrm{SST_{amp}}$ are large, whereas substantially negative $\mathrm{AI_{PW}}$ values are found for other cases. This result indicates that the ensemble does not sample a continuum of arbitrary rainbelt structures, but rather separates into a zonally extended regime and a zonally self-aggregated regime. ZCSA regime is preferentially realized under high-$\mathrm{SST_{max}}$ and high-$\mathrm{SST_{amp}}$ conditions.

This behavior is partly consistent with MH20: within the subset of cases that undergo ZCSA, larger $\mathrm{SST_{amp}}$ at fixed $\mathrm{SST_{max}}$ tends to delay its development (compare max305amp3 with max305amp5, and

max303amp3 with max303amp5), analogous to their result that a stronger SST contrast delays the onset of ZCSA. A notable difference is that, in our simulations, $SST_{max}$ and $SST_{amp}$ influence not only the timing of ZCSA but also its occurrence itself over the explored parameter space, whereas ZCSA developed in all three SST-contrast experiments they considered in MH20. This broader sensitivity shifts the emphasis of the present study toward the onset problem itself. In the following parts (Figs. 3 onward), we therefore investigate the physical mechanisms that distinguish cases in which ZCSA emerges from those in which it does not.

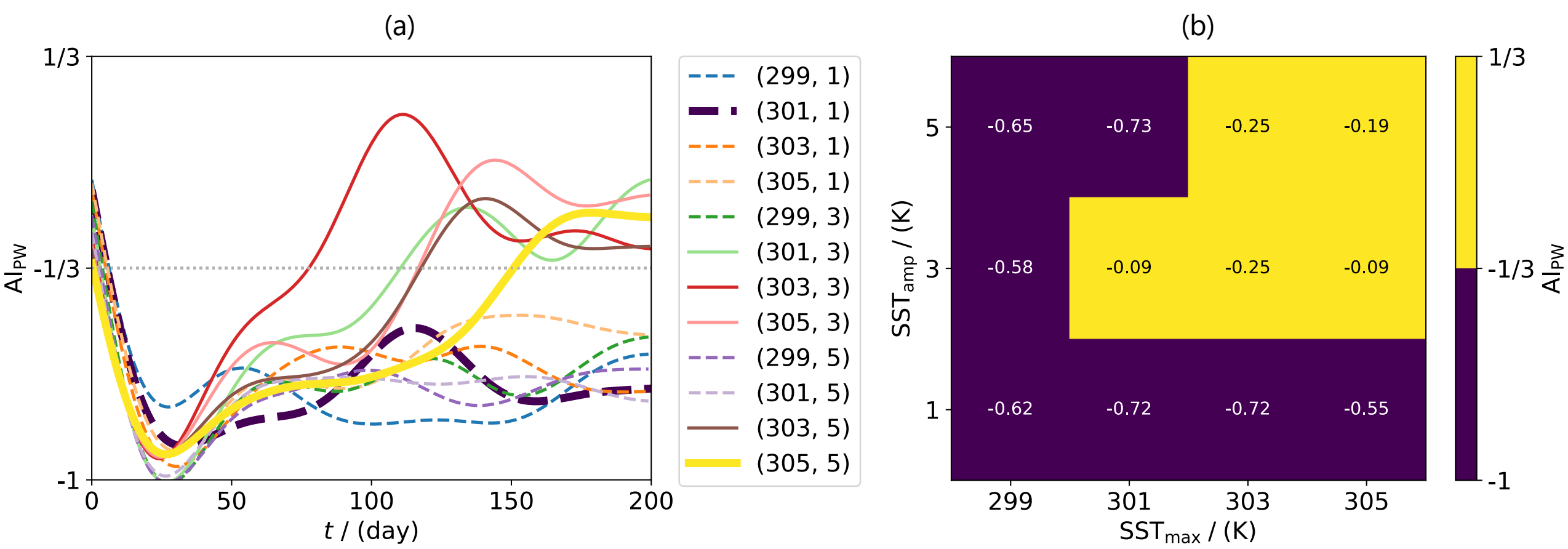

- Figure 2. (a) Time series of the PW anisotropy index, $AI_{PW}$, for all 12 simulations. Each line is labeled as ($SST_{max}$, $SST_{amp}$) format in the unit of K. The horizontal gray dotted line displays the regime boundary: $AI_{PW} = -1/3$. Each daily-mean time series is processed with a 60-day low-pass filter to focus on the slowly varying component. The two cases plotted by a bold line (max301amp1 and max305amp5) correspond to the representative cases shown in Fig.1. (b) Parameter-space distribution of the time mean $AI_{PW}$ over the final 30 days of each 200-day simulation as a function of $SST_{max}$ and $SST_{amp}$. Text labels indicate the actual $AI_{PW}$ values.

The temporal evolution of a representative ZCSA case, max305amp5, is shown in Fig. 3. In the early stage of the time integration (days 1–30), moisture and low-level meridional convergence develop primarily over the warm central part of the domain, indicating the formation of an initially zonally extended rainbelt. The zonally averaged fields (Figs. 3a and c) show that this early rainbelt is accompanied by strong meridional surface inflow from the cooler sides toward the high-SST region and by a pronounced meridional contrast in PW.

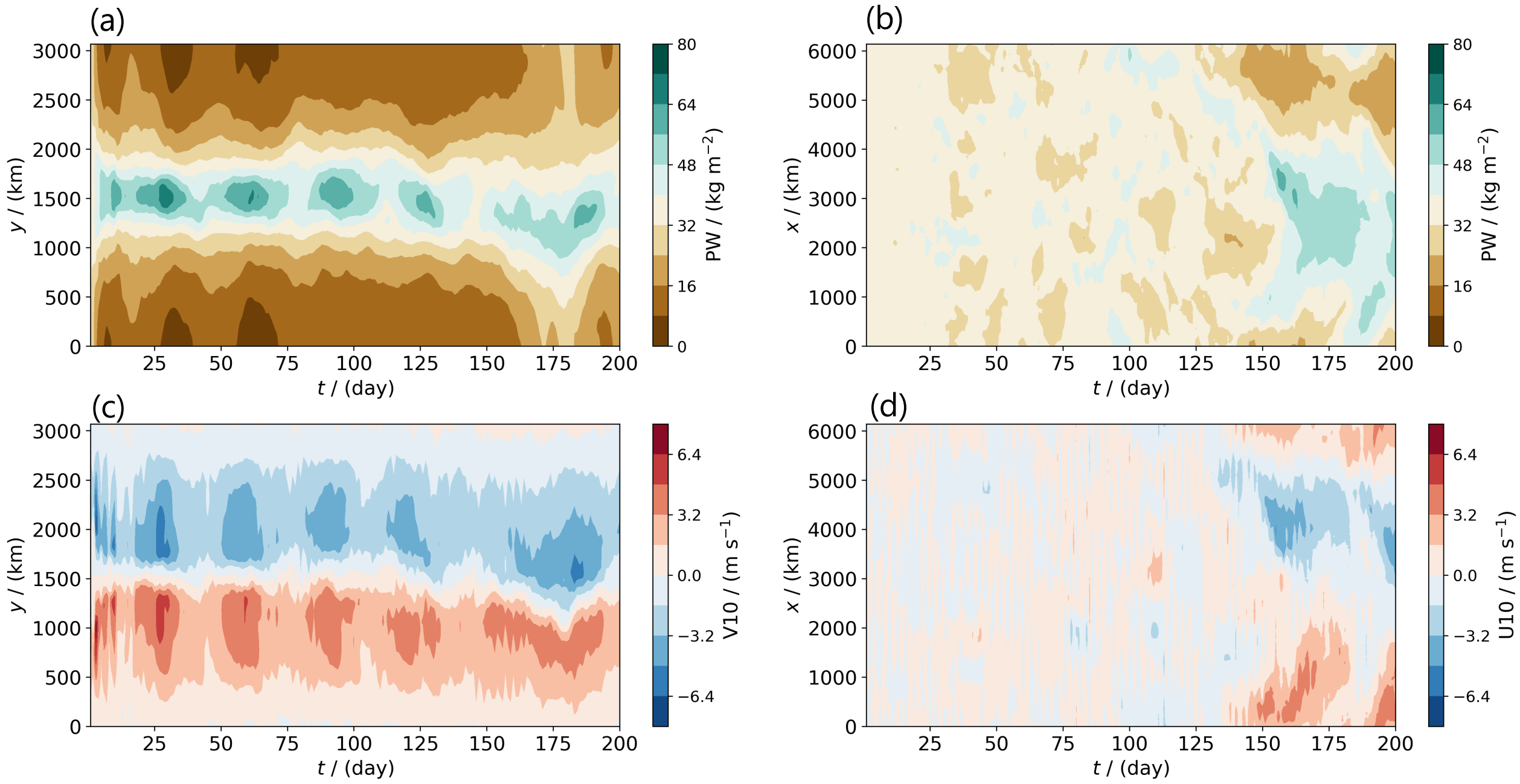

- Figure 3. Hovmöller diagrams for the representative ZCSA case max305amp5. (a) Zonally averaged PW as a function of meridional position and time. . (b) Meridionally averaged PW in the high-SST region as a function of zonal position and time. (c) Zonally averaged meridional surface wind as a function of meridional position and time. (d) Meridionally averaged zonal surface wind in the high-SST region as a function of zonal position and time. The high-SST region is defined as the subdomain where SST is higher than $SST_{ave}$. All values shown here are daily means.

As the simulation evolves, however, the system does not remain in a steady meridionally organized state. Instead, it exhibits temporal fluctuations between phases of relatively strong and relatively weak meridional surface convergence (days 31–135). These phases are accompanied by corresponding changes in the meridional moisture contrast: periods of strong meridional inflow are associated with larger meridional PW differences, whereas periods of weakened meridional inflow coincide with a reduced meridional moisture contrast.

A key feature is that the development of zonal structure (Figs. 3b and d) occurs after the system enters one of these weak meridional convergence phases around days 125–150 (Fig. 3a). In the later stage of the simulation, the meridionally averaged fields over the high-SST region (Figs. 3b and d) show the emergence of pronounced zonal moisture inhomogeneity together with zonal surface circulation. Thus, in this case, ZCSA does not arise directly from the initially formed rainbelt, but develops after a temporary weakening of the meridional low-level convergence that had maintained the zonally extended rainbelt. In the next subsection, we examine how general this feature is across the parameter space and what controls the weakening of the meridional surface flow.

### 3.2 BL control of the weakened meridional circulation

We now examine the weakening of the meridional near-surface circulation more systematically across the parameter space. Figure 4a summarizes, for each simulation, the temporal minimum of the zonally-averaged daily-mean meridional surface convergence after the initial spin-up period (avoiding days 1–20 in detecting temporal minimum). Here the meridional surface convergence $\Delta v_{\mathrm{sfc}}$ is defined as the difference between the meridional maximum and minimum of zonally-averaged daily-mean meridional surface wind, and each time series is 10-day low-pass filtered to emphasize the slowly varying component of the circulation. Figure 4 shows that all ZCSA cases exhibit a distinct phase of weakened meridional surface convergence. This indicates that the temporary muting of the meridional near-surface circulation is not unique to the representative case (max305amp5 shown in Fig. 3c), but is a robust feature of the simulations that later develop ZCSA.

At the same time, Fig. 4a also shows that weak meridional convergence is not by itself a sufficient condition for ZCSA. In particular, the cases max303amp1 and max305amp1, which do not develop ZCSA (Fig. 2b), also exhibit relatively weak meridional surface convergence (Fig. 4a). This suggests that the onset of ZCSA involves at least two distinct ingredients: first, a temporary weakening of the mean meridional circulation that relaxes the zonally elongated rainbelt structure, and second, an additional condition that determines whether zonal organization can subsequently grow. We postpone the second question to the energetic analysis in Section 3.3, and focus here on the first: what controls the weakening of the meridional surface flow from a BL momentum perspective?

First, it is useful to ask whether the weaker meridional surface convergence could be explained simply by weaker meridional surface pressure difference. Figure 4b addresses this point using the temporal minimum of the surface-pressure difference between the high-SST and low-SST regions, $\Delta p_{\mathrm{sfc}} = p_{\mathrm{sfc,lowSST}} - p_{\mathrm{sfc,highSST}}$, where $p_{\mathrm{sfc,lowSST}}$ and $p_{\mathrm{sfc,lowSST}}$ are the horizontal means of surface pressure over the subdomains with SST above and below $\mathrm{SST}_{\mathrm{ave}}$, respectively. Because $\Delta p_{\mathrm{sfc}}$ is positive in all simulations, its temporal minimum corresponds to the weakest pressure contrast. Figure 4b shows that simulations with weaker meridional surface convergence (Fig. 4a) do not systematically have a weaker pressure contrast. Thus, the weak-convergence state cannot be explained simply by weaker pressure-gradient forcing. This point is also illustrated by comparing the representative cases max301amp1 and max305amp5: despite weaker meridional convergence $\Delta v_{\mathrm{sfc}}$ happens in the latter, its minimum $\Delta p_{\mathrm{sfc}}$ is not smaller. This suggests that the weak-wind state arises not simply from weaker driving, but from stronger effective damping.

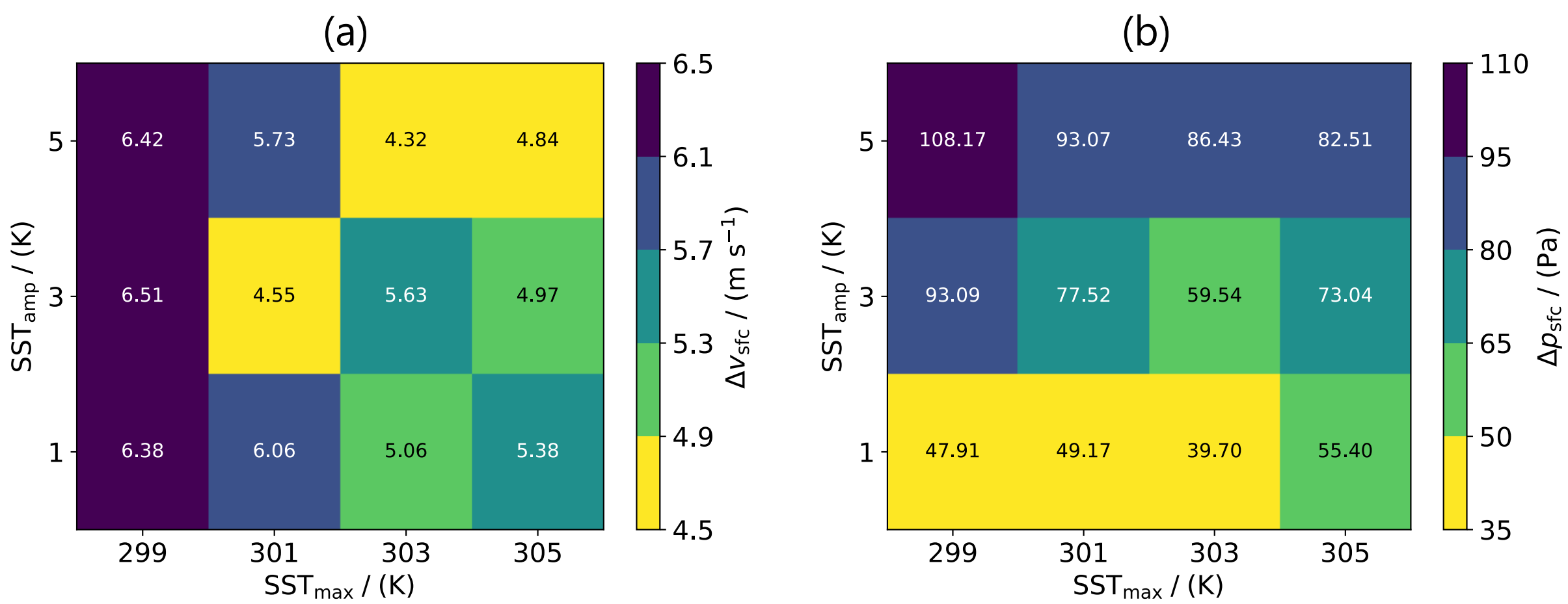

- Figure 4. Temporal weakening of meridional surface convergence and meridional surface-pressure difference across the $\mathrm{SST_{max}}$–$\mathrm{SST_{amp}}$ parameter space. (a) Temporal minimum of $\Delta v_{\mathrm{sfc}}$; $\Delta v_{\mathrm{sfc}}$ is defined as the difference between the meridional maximum and minimum of zonally-averaged meridional surface wind. (b) Temporal minimum of $\Delta p_{\mathrm{sfc}}$; $\Delta p_{\mathrm{sfc}}$ is the surface pressure difference between the high-SST and low-SST regions (see main text). Because $\Delta p_{\mathrm{sfc}}$ is positive in all simulations, its temporal minimum corresponds to the weakest surface-pressure contrast between the two regions. Each time series is processed with a 10-day low-pass filter to emphasize the slowly varying fields. The initial 20 days are excluded when detecting the minimum, because they correspond to the initial development stage of the circulation. In (a), weaker convergence is colored in more yellowish to see whether it appears within the same parameter range as ZCSA shown in Figure 2a. In (b), weaker pressure difference is colored in yellowish to see whether it appeas within the same parameter range as weaker convergence shown in (a).

To clarify this point, we next examine the BL-averaged meridional momentum budget in a representative non-ZCSA case (max301amp1) and a representative ZCSA case (max305amp5) (Fig. 5). In this study, BL depth is defined as the altitude at which the virtual potential temperature first exceeds that of first layer.

Each term is expressed as an effective rate (in a unit of inverse time), defined as the budget tendency normalized by the BL meridional wind, and the time series are 10-day low-pass filtered to emphasize the slowly varying component of the circulation. The details of the analysis are provided in the Supplementary Material S1.

In both simulations, during the earlier adjustment stage (days 1–30), the meridional pressure-gradient-force (PGF), surface friction (SFC), and the residual (RES) terms are comparably important, whereas the advection (ADV) term associated with the zonally-averaged circulation remains relatively small. Here, the residual term includes the effect of entrainment at the BL top. Once the large-scale circulation is established (around days 31–50), the dominant balance is between acceleration by the PGF and deceleration by SFC. In the non-ZCSA case, the PGF-driven acceleration is balanced by a moderate surface-friction damping, allowing a relatively strong meridional BL flow to persist (> 2 m $s^{-1}$). In the ZCSA case, by contrast, the BL meridional wind remains substantially weaker, and this weak-wind state (< 2 m $s^{-1}$) is associated with a stronger effective damping by surface friction.

Here the effective damping should not be interpreted as a prescribed linear drag coefficient. Rather, it is a diagnosed, state-dependent quantity that reflects how strongly surface drag and related BL processes act on the BL-mean momentum (Large and Pond 1981; Stevens 2006). In standard tropical mixed-layer formulations (Stevens et al. 2002; Back and Bretherton 2009b), frictional and entrainment-related tendencies act on the BL flow with a strength that depends on BL depth, so the same meridional PGF need not produce the same wind response in different cases. This can be illustrated by the approximate balance for the zonally averaged BL meridional wind between PGF and SFC at a given location,

$$0 = -\partial_y \hat{p} + \frac{\tau_0}{H_{\mathrm{BL}}},$$

where $\hat{p}$ is the BL averaged pressure, $\tau_0$ the surface stress, and $H_{\mathrm{BL}}$ the BL depth (see Supplementary Material S1 for details). The key point is that, for a given surface stress, the momentum sink imposed at the surface is distributed over the depth of the BL. When the BL is shallower, the same surface momentum loss has a larger impact on the BL-mean momentum and therefore produces a larger effective damping rate.

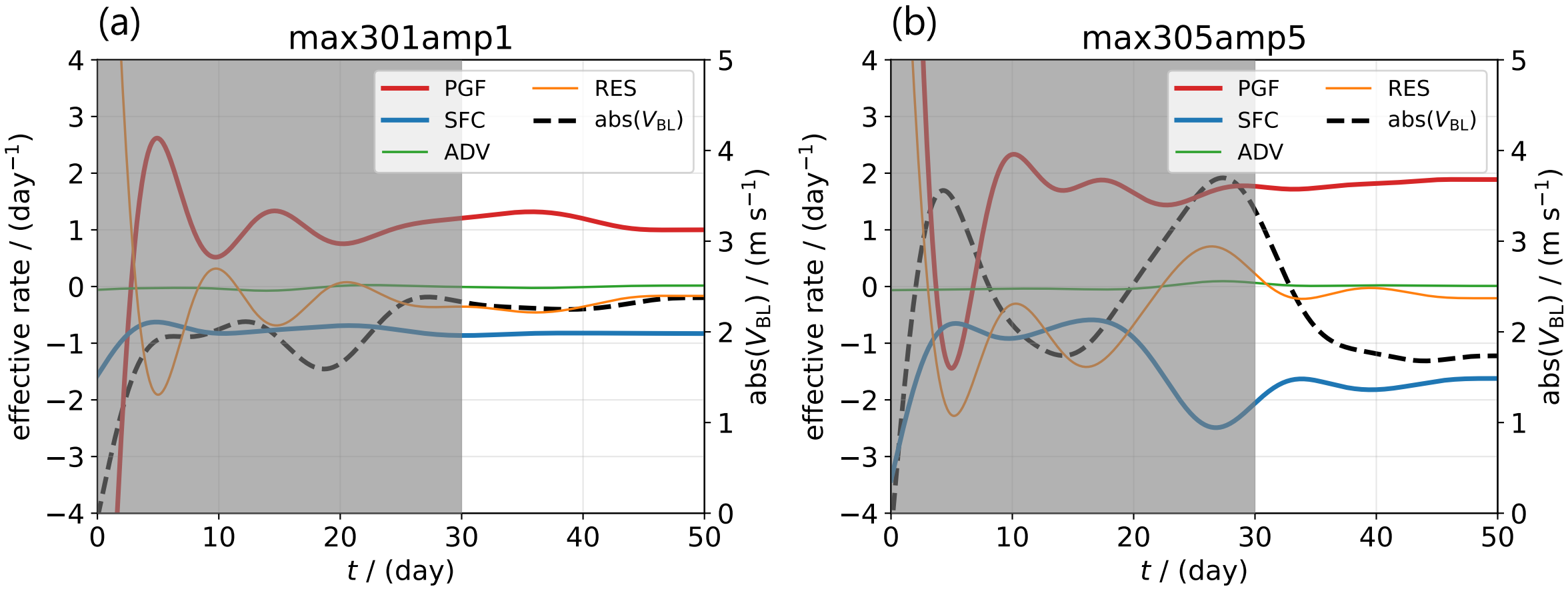

- Figure 5. Time series of BL-averaged meridional momentum budget terms for (a) max301amp1 and (b) max305amp5. Each term is shown as a domain-mean effective rate, defined as the budget tendency normalized by the BL meridional wind. PGF denotes the pressure-gradient-force term, SFC the surface-friction term, ADV the advection term associated with the zonally averaged circulation, and RES the residual term including the effect of entrainment at the BL top. The mean magnitude of the BL-averaged meridional wind is shown on the right vertical axis in each panel. All time series are processed with a 10-day low-pass filter. The details are provided in Supplement S1.

This interpretation can be tested by examining the BL depth itself. As shown in Fig. 6a, the representative ZCSA case (max305amp5) has a shallower BL than the non-ZCSA case (max301amp1). At the same time, max305amp5 exhibits a stronger meridional pressure difference but a weaker meridional inflow (Fig. 4). This combination is consistent with the effective-damping interpretation discussed above: when the BL is shallower, a comparable surface momentum sink acts on a smaller column of BL momentum and thus produces a larger damping rate. Moreover, Fig. 6a indicates that cases that tended to favor ZSCA are also cases with comparatively shallow BL.

We therefore next ask why the BL becomes shallow, and which environmental factor best explains its variation across the parameter space. In a mixed-layer framework, BL depth is primarily controlled by lower-tropospheric stability (LTS), radiatively driven subsidence, and surface buoyancy flux (Lilly 1968). A shallower BL is generally favored by larger LTS, stronger radiatively driven subsidence, and weaker surface buoyancy, whereas the opposite conditions favor a deeper BL.

Figure 6b–d summarize these candidate controls of BL depth over the parameter space during days 21–80, that is, after the initial spin-up but before fully transitioning to the ZCSA regime (Fig. 2a). A clear

relationship emerges: cases with a shallower BL in the low-SST region are generally those with larger LTS. This tendency is especially evident in the part of the parameter space where ZCSA occurs. By contrast, neither the radiative subsidence nor the surface buoyancy flux shows variations that explain the shallow BL as systematically as LTS does. These results support the interpretation that enhanced LTS is the primary control on BL depth in the present set of experiments. Together with the momentum-budget analysis, these results suggest that the weak meridional inflow in the ZCSA-favorable cases is tied to enhanced LTS: larger LTS favors a shallower BL, and a shallower BL makes surface drag act more effectively on the BL-mean momentum.

The dependence of LTS on $\mathrm{SST_{max}}$ and $\mathrm{SST_{amp}}$ is also physically plausible from the imposed thermodynamic structure. For a fixed $\mathrm{SST_{amp}}$, increasing $\mathrm{SST_{max}}$ tends to warm the free troposphere through moist-adiabatic adjustment over the warm convecting region. Under the weak temperature gradient (WTG) conditions in the present non-rotating simulations, this warming is communicated broadly across the domain, leading to a domain-wide increase in LTS. By contrast, for a fixed $\mathrm{SST_{max}}$, increasing $\mathrm{SST_{amp}}$ lowers $\mathrm{SST_{ave}}$ and tends to cool the BL, especially over the low-SST region where near-surface temperature is more directly tied to the local SST. Since the free tropospheric temepratature is largely constrained by the fixed $\mathrm{SST_{max}}$ through WTG adjustment, a cooler BL increases LTS.

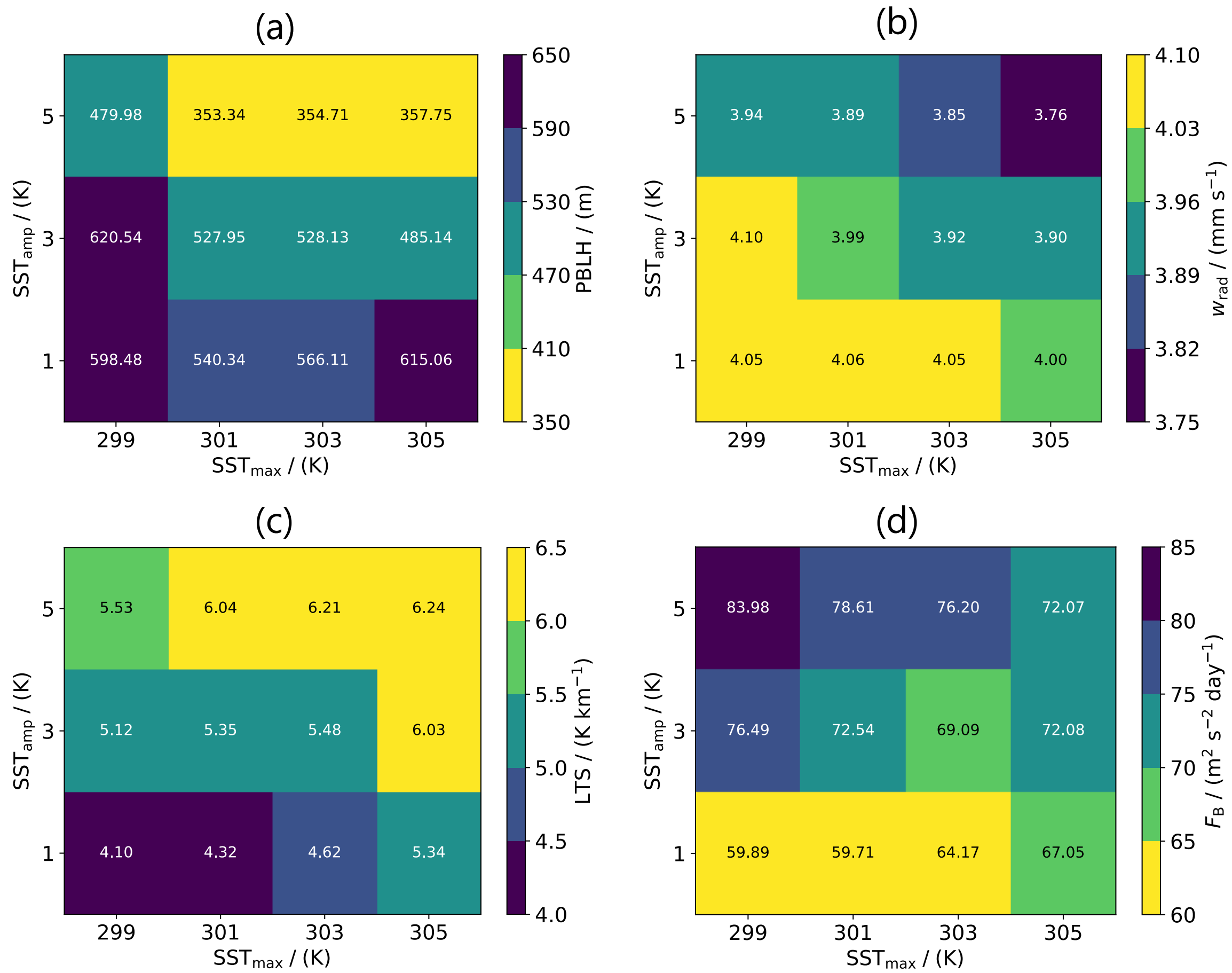

- Figure 6. Time-mean values of BL depth and its controlling factors. (a) Domain-averaged BL depth. (b) Domain-averaged radiatively driven vertical velocity, defined as the column radiative cooling divided by the static stability, $\left(\mathrm{d}T/\mathrm{d}z + g/c_p\right)$, averaged over the layer between the surface and the tropopause (minimum temperature) height. (c) Domain-averaged LTS, defined as the static stability averaged over the layer between the surface and 2.5 km height. (d) Domain-averaged surface buoyancy flux. All quantities are averaged over days 21–80. In (a), shallower BL depth is colored in more yellowish to see whether it appears within the same parameter range as weaker convergence in Figure 4. In (b)–(d), stronger radiative subsidence, larger LTS, and weaker surface buoyancy flux are colored in more yellowish to see whether they appear within the same parameter range as shallower BL depth in (a).

### 3.3 Energetic condition and transport reorganization associated with ZCSA

The BL analysis above explains how the meridional low-level circulation can become temporarily weak, but it does not by itself explain why such weakened-convergence states lead to ZCSA only in some cases. This distinction suggests that weak meridional convergence is a necessary enabling condition, but not a sufficient one. We therefore examine the problem from an energetic perspective and ask whether cases that develop ZCSA are also characterized by a larger imposed meridional contrast in MSE forcing.

Figure 7 shows the difference in MSE forcing between the high-SST and low-SST regions, where the high-SST and low-SST regions are defined as the subdomains with SST higher and lower than $\mathrm{SST_{ave}}$, respectively. Here the MSE forcing is defined as the sum of the surface sensible heat flux, surface latent heat flux, and column radiative heating. Under statistical equilibrium, the circulation must export from each column the MSE supplied by these forcings, so the meridional difference in MSE forcing provides a simple proxy for the large-scale energetic contrast imposed between the warm and cool parts of the domain. The physical background of this MSE analysis is summarized in Supplementary Material S2.

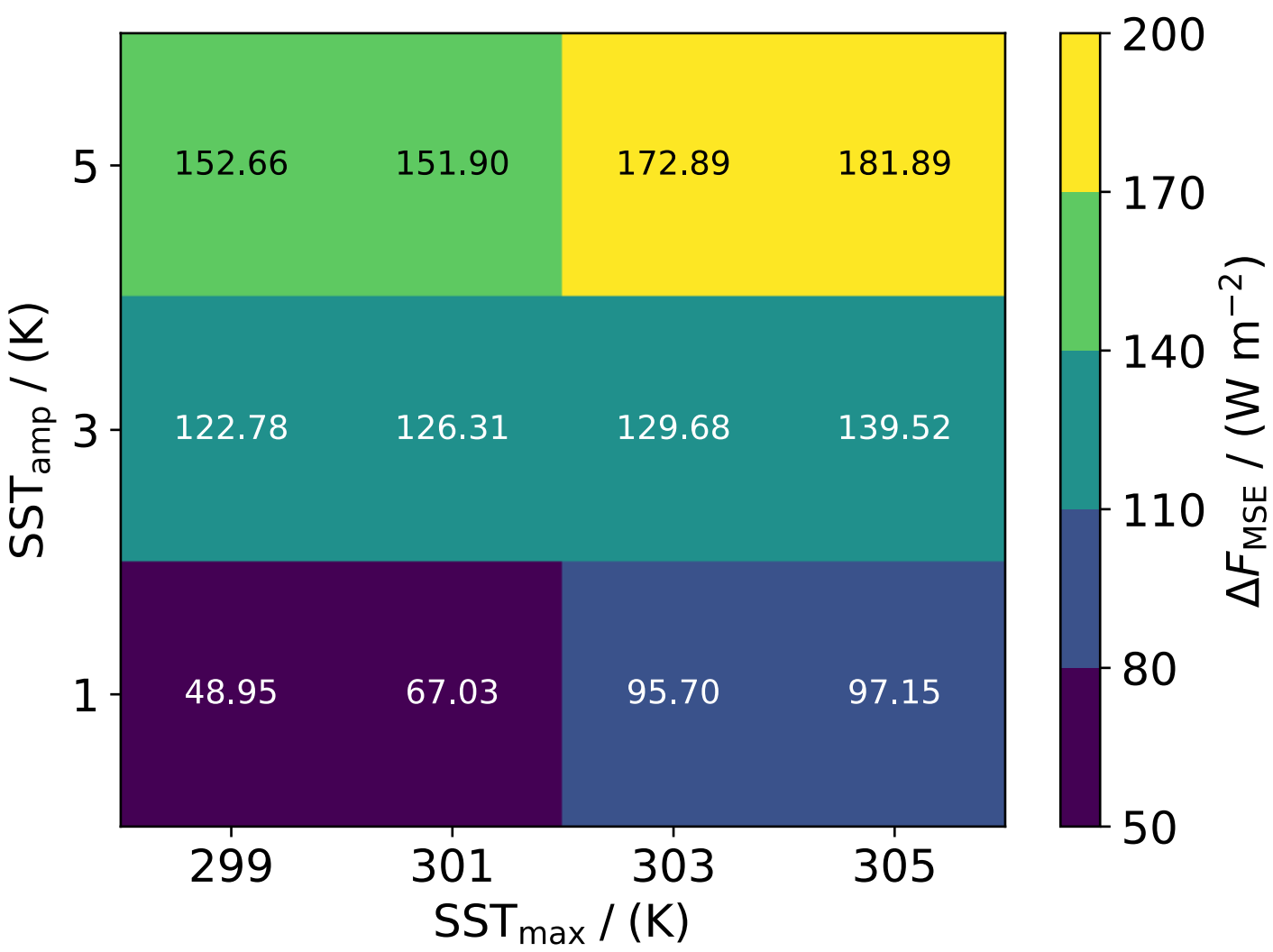

- Figure 7. The difference in MSE forcing averaged over high-SST regions and low-SST regions, $\Delta F_{\mathrm{MSE}}$. The MSE forcing $F_{\mathrm{MSE}}$ is defined as the sum of the surface sensible heat flux, surface latent heat flux, and column radiative heating. Positive value indicates that MSE forcing is larger in high-SST region than low-SST region. All quantities are averaged over days 21–80. The high-SST and low-SST regions are the subdomains where SST is higher and lower than $\mathrm{SST_{ave}}$, respectively.

A clear dependence on the imposed SST parameters is seen. The MSE-forcing difference $\Delta F_{\mathrm{MSE}}$ increases primarily with $\mathrm{SST_{amp}}$ and secondarily with $\mathrm{SST_{max}}$, indicating that the energetic contrast between the warm and cool parts of the domain becomes larger. In this sense, cases with high $\mathrm{SST_{amp}}$ and high $\mathrm{SST_{max}}$ place a stronger demand on the circulation to export MSE meridionally from the high-SST region to the low-SST region in order to maintain statictical equilibrium.

A larger MSE-forcing contrast is physically expected as $\mathrm{SST_{amp}}$ increases, although its quantitative magnitude is not trivial because the forcing includes interactive radiation as well as turbulent surface-flux contributions. A qualitative interpretation is that low-level air originating from the cool, dry low-SST region is advected into the high-SST region, where the air–sea humidity disequilibrium becomes large and favors strong evaporation. This tendency is expected to become more pronounced as $\mathrm{SST_{amp}}$ increases, because the inflowing air is colder and drier relative to the warm surface, and also as $\mathrm{SST_{max}}$ increases, because the saturation humidity over the warm SST rises rapidly with temperature. The diagnosed contrast is consistent with this picture and further shows that its increase is dominated by the latent heat flux term. For example, the MSE-forcing contrast increases from 67 $\mathrm{W\ m^{-2}}$ in max301amp1 to 182 $\mathrm{W\ m^{-2}}$ in max305amp5, with most of the change coming from the latent-heat-flux contrast (32 to 131 $\mathrm{W\ m^{-2}}$), while the sensible-heat-flux and radiative contributions change much less.

This result provides a possible explanation for why weak meridional convergence alone is not sufficient for ZCSA. Some cases with weakened meridional inflow, such as max303amp1 and max305amp1, remains characterized by relatively modest meridional MSE-forcing contrast, whereas the ZCSA cases combine weakened meridional low-level convergence (Fig. 4) with a large meridional energetic contrast (Fig. 7). This suggests that ZCSA tends to occur when the rainbelt is not only dynamically released from strong meridional low-level constraint, but also subject to a large high-SST-to-low-SST MSE transport requirement. At the same time, the present analysis does not demonstrate that this energetic contrast directly triggers zonal aggregation. Rather, it identifies a large-scale constraint that appears to distinguish many ZCSA and non-ZCSA cases.

To clarify how this constraint is accommodated once ZCSA develops, we next diagnose how the meridional MSE transport is realized and how its decomposition differs between ZCSA and non-ZCSA cases. Following the standard decomposition of atmospheric meridional transport into mean, stationary eddy, and transient eddy contributions (Lorenz 1967; Oort 1971), Fig. 8 compares the transport between the high-SST and low-SST regions for three representative states: the later phase (days 171–200) of the non-ZCSA case max301amp1, and the early phase (days 21–50) and the later phase (days 171–200) of the ZCSA case max305amp5. The early phase of max305amp5 is shown to see the characteristics before ZCSA becomes prominent. Positive values indicate net export from the high-SST region to the low-SST region.

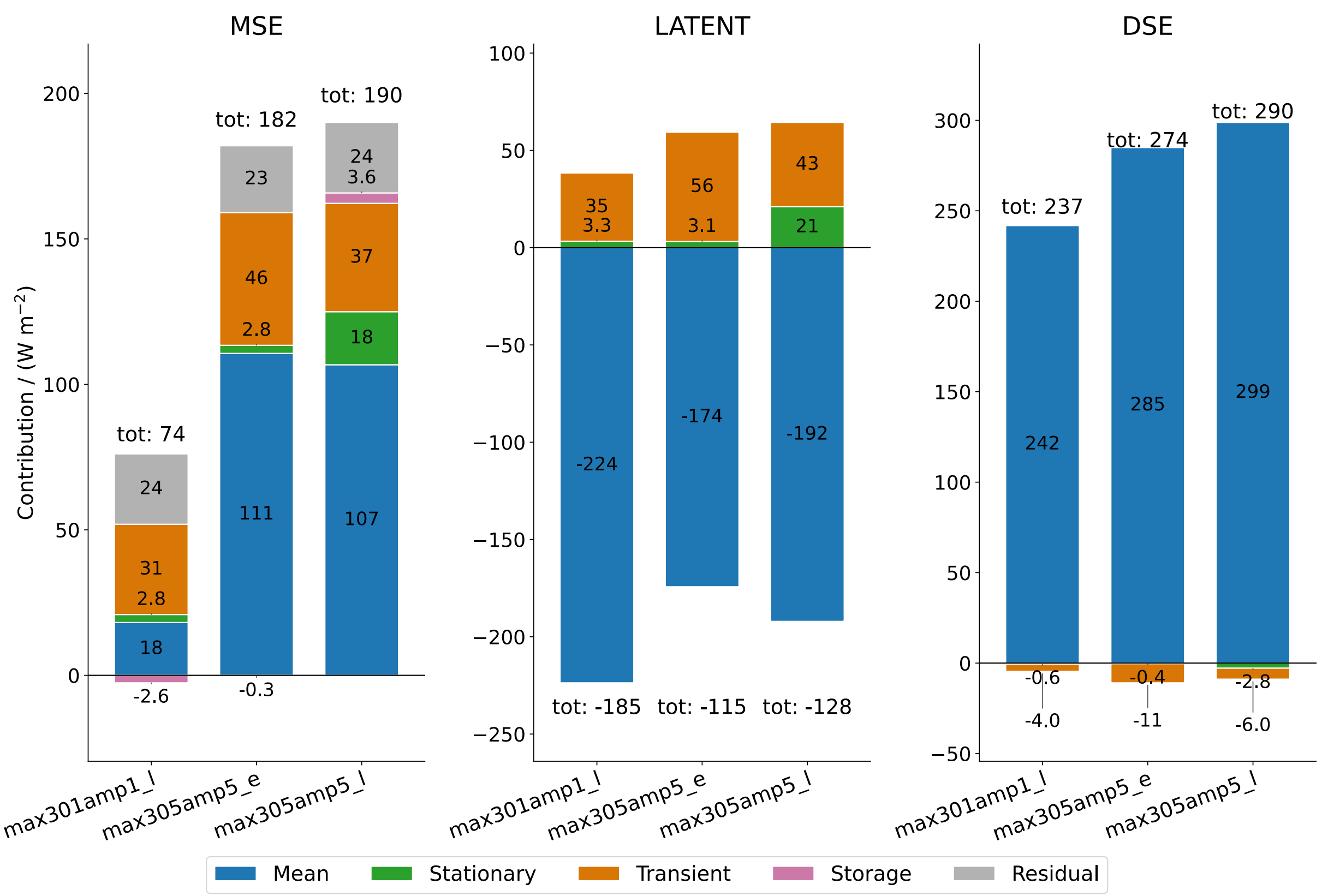

- Figure 8. Meridional MSE transport between the high-SST and low-SST regions. Left: decomposition of the total MSE transport into mean, stationary, and transient contributions, together with the storage term and the residual term. Positive values indicate export from the high-SST region to the low-SST region; for the storage term, positive values indicate an increase in the contrast between the two regions. Center and right: corresponding decompositions for the meridional transport of the latent energy (LATENT) and dry-static-energy (DSE) components. LATENT and DSE include only resolved transport contributions. Hence, the total MSE contribution does not necessarily equal the sum of the total LATENT and DSE contributions. Each panel shows three states: the later phase of the non-ZCSA case (max301amp1_l), the early phase of the ZCSA case (max305amp5_e), and the later phase of the ZCSA case (max305amp5_l). The early and later phases are defined as days 21–50 and days 171–200, respectively.

A clear reorganization of the transport occurs as the ZCSA case evolves. In the later phase of max305amp5, the export of total MSE is enhanced relative to both the non-ZCSA case and the early phase of the ZCSA case. This enhancement is associated in particular with an increase in the stationary eddy contribution, indicating that the persistent zonal structure that emerges in the mature ZCSA state contributes to the meridional redistribution of MSE. The increase in stationary eddy export is especially pronounced for the latent-energy component. At the same time, the total transport also changes in its mean component. In the later phase of the ZCSA case, the total latent-energy import and the total dry-static-energy (DSE) export are both increased, with the mean circulation accounting for most of these changes.

This interpretation is broadly consistent with MH20, who showed that zonal self-aggregation is accompanied by an intensified meridional moisture outflow and moistening of the cooler environment. The present analysis extends that picture by placing the transport reorganization in the context of the full MSE budget and by quantifying how much of the mature export is carried by the stationary component. A plausible interpretation is that, once the meridional low-level inflow is weakened, a zonally extended rainbelt becomes difficult to maintain as a broad convergence line, even though a substantial high-SST-to-low-SST MSE export is still required. Under such conditions, the system may reorganize into a zonally localized convective area. In that state, enhanced zonal convergence into the localized area can help sustain concentrated ascent and convective heating, and the resulting three-dimensional circulation may in turn support a stronger meridional export of moisture and MSE. In this view, zonal organization does not directly replace the required meridional transport; rather, it helps establish a circulation–moisture structure through which that transport can be maintained more robustly. A related possibility is that the larger-scale zonal moisture structure in the mature ZCSA state is less susceptible to lateral dilution than a more weakly organized state, helping the stationary transport persist once established. Although the present analysis does not fully resolve the causal pathway and the whole picture of the interactions, it suggests a concrete working hypothesis: ZCSA emerges when a rainbelt that has been dynamically loosened by weak meridional inflow is simultaneously subject to a sufficiently strong meridional transport requirement.

## 4. Discussion

A central implication of this study is that zonal symmetry breaking of an ITCZ-like rain belt can be understood as an emergent moist-dynamical instability. In the present simulations, ZCSA is not simply cloud clustering superimposed on a fixed background state. Rather, it is a transition of the rain belt itself from a zonally extended structure to a zonally localized one, and this transition depends on the coupled evolution of thermodynamic structure, BL circulation, and meridional energetic contrast. The results support a two-condition picture. A first requirement is a temporary weakening of the meridional low-level circulation. In the ZCSA cases, this weakening is associated with larger LTS over the cool subsiding region, which leads to a shallower BL and makes surface drag more effective in damping the meridional inflow. However, this is not sufficient by itself, because some non-ZCSA cases also pass through weak-convergence states. A second requirement is therefore a sufficiently strong meridional energetic contrast.

When the MSE forcing contrast between the warm and cool regions is large, the circulation must still export substantial energy meridionally, and ZCSA appears as one way in which this transport can be reorganized.

The mechanism identified here offers a useful perspective on tropical phenomena in which rain belts lose zonal coherence or develop weak-wind regions. In particular, the present results suggest that spontaneous zonal organization under zonally uniform lower-boundary conditions may represent one fundamental pathway by which an ITCZ-like rain belt becomes unstable to breakdown. ITCZ breakdown is a well-known feature of the tropical atmosphere, and previous interpretations have often emphasized wave, eddy, or vortex-dynamical aspects of the problem (e.g., Ferreira and Schubert 1997). More recently, observed breakdown events have also been linked to anomalous moisture advection associated with equatorial Rossby-wave activity (Casallas et al. 2026). The present results are complementary to those views. Rather than focusing on externally triggered synoptic disturbances, they show that an ITCZ-like rain belt can possess an intrinsic tendency toward zonal symmetry breaking when the meridional low-level flow is weakened and the circulation remains energetically stressed. We therefore interpret our mechanism not as a complete explanation of observed ITCZ breakdown, but as a possible foundational ingredient that may interact with wave and vortex dynamics in nature. Another suggestive implication concerns weak-wind regions in the tropics. In the ZCSA cases, zonal organization is associated with weak near-surface winds over the warm part of the domain. This is loosely reminiscent of the observed coexistence of low winds and warm ITCZ environments in the tropical Atlantic (Windmiller 2024). Because the present simulations omit rotation, seasonal migration, and the full trade-wind structure, they are not intended as a direct model of the real doldrums. Rather, they suggest more generally that weak-wind tropical regimes can arise from the coupling between moist convection, BL structure, and rain-belt organization itself.

Furthermore, it is noteworthy that the evolution of moisture variability in the ZCSA cases also suggests a broader connection to recent conceptual ideas on moisture-mode–Hadley interactions. In the representative ZCSA case, the meridional moisture contrast grows during the early stage of rain-belt formation, but later declines while zonal moisture variability continues to increase (Figs. 2a, 3a and 3b); this transition is not seen in the same way in the non-ZCSA case. This behavior is qualitatively consistent with the picture proposed by Adames and Mayta (2024), in which moisture-mode activity grows at the expense of a zonal-mean background meridional moisture gradient. Our diagnostics are not a formal test of that theory, but they suggest that ZCSA may provide a useful bridge between explicit-convection modeling of rain-belt organization and recent theoretical ideas on the coupling between tropical wave activity and mean overturning—Hadley circulation.

As limitations of the present work, part of the mechanism acts through BL depth and effective frictional damping, both of which may depend on model resolution and turbulent transport. In particular, the quantitative relationship between LTS, BL depth, and effective drag should be re-examined in turbulence-resolving simulations. However, the dependence of LTS on the imposed SST pattern arises from large-scale thermodynamic constraints—most notably the moist adiabat over the convecting region and its remote influence through WTG adjustment—and is therefore likely to be qualitatively robust even if the detailed BL response changes with resolution. Another important limitation is that, although the present analysis identifies a large meridional MSE-forcing contrast as a characteristic of the ZCSA-favorable cases, it does not yet clarify through what physical mechanism this energetic constraint contributes to the onset and selection of zonal aggregation. The transport decomposition shows how the mature ZCSA state accommodates the required export, but the causal linkage between the imposed energetic contrast and the growth of zonal structure remains an important subject for future work.

## 5. Conclusions

This study examined the zonal symmetry breaking of an idealized rain belt under a prescribed meridional SST distribution by varying the peak SST ($\mathrm{SST_{max}}$) and the meridional SST amplitude ($\mathrm{SST_{amp}}$). The main conclusions are as follows.

First, the simulations separate into two structural regimes: a zonally elongated regime, in which the rain belt remains zonally homogeneous, and a zonally aggregated (or clustered) regime, in which pronounced zonal moisture inhomogeneity develops. The latter regime, zonal convective self-aggregation (ZCSA), occurs preferentially when both $\mathrm{SST_{max}}$ and $\mathrm{SST_{amp}}$ are large.

Second, all ZCSA cases pass through a phase of weakened meridional near-surface convergence. This weak-convergence phase appears to be a necessary enabling condition for ZCSA, although it is not sufficient by itself.

Third, the weakened meridional inflow is linked to stronger effective frictional damping in the boundary layer (BL). This enhanced damping is associated with a shallower BL over the cool subsiding region, which is most consistently explained by larger lower troposhperic stability (LTS). In the present configuration, $\mathrm{SST_{max}}$ and $\mathrm{SST_{amp}}$ both contribute to this LTS increase by, respectively, warming the free troposphere through moist-adiabatic adjustment and cooling the BL over the low-SST region.

Fourth, weak meridional convergence alone does not determine whether ZCSA occurs. The ZCSA cases are additionally characterized by a large meridional contrast in moist static energy (MSE) forcing, implying a strong requirement for meridional MSE transport. Thus, ZCSA is favored when a weakened meridional low-level constraint coexists with a strong energetic contrast between the warm and cool regions.

Finally, ZCSA is accompanied by a systematic reorganization of both transport and moisture structure. The meridional MSE transport exhibits enhanced stationary eddy export from the warm region, and the moisture field shifts from a predominantly meridional contrast toward increased zonal variability.

Taken together, these results suggest that zonal symmetry breaking of an ITCZ-like rain belt is favored when a weakened meridional inflow, associated with shallow-BL damping, coincides with a large imposed meridional MSE-forcing contrast. More broadly, this study highlights the coupled roles of convective thermodynamics, weak-temperature-gradient coupling, boundary-layer momentum balance, and large-scale energetic constraints in tropical rain-belt organization.

## Acknowledgements

This study was partly supported by JSPS KAKENHI Grant JP24K17128. T.Y. was also supported by the Alexander von Humboldt Research Fellowship for a research stay at the Max Planck Institute for Meteorology, and by the RIKEN Special Postdoctoral Researcher Program for preliminary research at the RIKEN. This work used computational resources of Fugaku provided by RIKEN Center for Computational Science through the HPCI System Research Project (Project ID: hp250308). T.Y. thanks Hirofumi Tomita and Bjorn Stevens for insightful discussions on the experimental design and preliminary results at an early stage of this work, and Hans Segura and Divya Sri Praturi for helpful discussions on the energy-transport analysis. The authors also thank Divya Sri Praturi for helpful internal review comments on an earlier version of the manuscript at the Max Planck Institute for Meteorology. The authors are grateful to Team SCALE for developing and providing the numerical model SCALE-RM.

## Data Availability Statement

The source code of SCALE-RM is publicly available under the 2-Clause BSD license (https://scale.riken.jp/). The scripts and data for reproducing the figures used for this study are publicly available in the Open Science Framework (https://osf.io/kg8du/).

## Supplementary Material

### S1. Boundary-layer meridional momentum budget

To diagnose the boundary-layer control of the meridional circulation, we start from the zonally-averaged boundary-layer-mean meridional momentum budget. Here and below $[f] = L_x^{-1} \int_0^{L_x} f \, dx$ denotes the zonal mean of a arbitrary variable $f$.

Using the zonal-mean density $[\rho]$ and zonal-mean meridional wind $[v]$, we define the boundary-layer mass $W$ and boundary-layer-integrated meridional momentum $M$ as

$$W(y,t) = \int_0^{H_{\mathrm{BL}}} [\rho] \mathrm{d}z, \quad M(y,t) = \int_0^{H_{\mathrm{BL}}} [\rho][v] \mathrm{d}z,$$

where $H_{\mathrm{BL}} = H_{\mathrm{BL}}(y,t)$ is the zonal-mean diagnosed boundary-layer height, which varies with meridional position and time, and is defined as the altitude at which the virtual potential temperature first exceeds that of first layer.

The boundary-layer-mean meridional wind $V_{\mathrm{BL}}$ is then

$$V_{\mathrm{BL}}(y,t) = \frac{M}{W}.$$

Although the model itself is fully compressible, the present diagnosis is formulated in terms of zonally averaged fields, consistent with the focus on the large-scale boundary-layer circulation.

We diagnose the tendency of the boundary-layer-mean meridional wind normalized by $V_{\mathrm{BL}}$ as

$$\mathrm{TEND} \equiv \frac{\partial_t V_{\mathrm{BL}}}{V_{\mathrm{BL}}},$$

and decompose it into pressure-gradient, mean-flow advection, surface-stress, and residual terms,

$$\mathrm{TEND} = \mathrm{PGF} + \mathrm{ADV} + \mathrm{SFC} + \mathrm{RES}.$$

Here

$$\mathrm{PGF} = -\frac{1}{W V_{\mathrm{BL}}} \int_0^{H_{\mathrm{BL}}} \partial_y \, [p] \mathrm{d}z,$$

$$\mathrm{ADV} = -\frac{1}{W V_{\mathrm{BL}}} \int_0^{H_{\mathrm{BL}}} [\rho]([v] \, \partial_y [v] + [w] \, \partial_z [v]) \mathrm{d}z,$$

and

$$\mathrm{SFC} = \frac{\tau_0}{W V_{\mathrm{BL}}},$$

where $p$ is pressure, $w$ is vertical velocity, and $\tau_0$ is the surface stress. The surface stress generally acts opposite to the flow and therefore contributes to damping. The residual term RES is obtained as the remainder of the budget. It is expected to include primarily the effect of entrainment and momentum flux at the boundary-layer top, together with other contributions not explicitly separated in the present diagnosis, such as the zonal eddy term.

The normalization by $V_{\mathrm{BL}}$ is aimed to compare the relative importance of the budget terms independently of the wind amplitude, and it defines the effective rate of each term in a unit of inverse time. In the main text (Fig. 5), although the budget is first defined as a function of meridional position, the plotted quantities are reduced to domain-wide meridional averages in order to highlight the temporal evolution of the overall boundary-layer momentum balance.

If the dominant balance is between PGF and SFC, then

$$0 = \mathrm{PGF} + \mathrm{SFC},$$

and assuming the density in the boundary layer is constant $\rho_0$, and the boundary-layer average pressure is denoted as $\hat{p}$, the balance becomes

$$0 = -\frac{\partial_y \hat{p}}{\rho_0 V_{\mathrm{BL}}} + \frac{\tau_0}{\rho_0 H_{\mathrm{BL}} V_{\mathrm{BL}}}.$$

This is a simplified version of mixed layer momentum balance model shown in Stevens et al. (2002) and Back and Bretherton (2009b); more specifically, excluding their entrainment and Coriolis term.

## S2. Meridional MSE-forcing imbalance and decomposition of meridional energy transport

To relate zonal convective self-aggregation to the large-scale energy budget, we consider the difference in zonally-averaged, vertically-integrated moist energy between the high-SST and low-SST regions. The high-SST region is defined as the half of the domain where the prescribed SST is larger than its domain-mean value $\mathrm{SST}_{\mathrm{ave}}$, and the low-SST region is defined as the other half where the prescribed SST is smaller than $\mathrm{SST}_{\mathrm{ave}}$.

For the transport analysis, the relevant scalar is moist static energy (MSE),

$$h = C_p T + gz + Lq,$$

where $C_p$ is specific heat at constant pressure, $T$ temperature, $g$ gravitational acceleration, $z$ height, $L$ the latent heat of vaporization, and $q$ specific humidity. In the decomposition shown in Fig. 9 of the main text, we further define latent energy and dry static energy (DSE) as

$$\text{LATENT} = Lq,$$

$$\text{DSE} = C_p T + gz,$$

so that

$$\text{MSE} = \text{LATENT} + \text{DSE}.$$

Following the standard Reynolds-type decomposition of meridional transport, the zonal-mean, time-mean meridional MSE transport is decomposed into mean, stationary eddy, and transient eddy contributions as

$$[\overline{\rho v h}] = [\,\overline{\rho v}\,]\,[\,\overline{h}\,] + [\,(\,\overline{\rho v}\,)^*\,\overline{h}^*\,] + [\,\overline{(\rho v)' h'}\,],$$

where $[f]$ denotes the zonal mean, the overbar $\bar{f}$ denotes a time mean over the analysis window, $\overline{f}^* = \overline{f} - [\overline{f}]$ is the zonal deviation of the time-mean field at a point with coordinate $x$, and $f' = f - \overline{f}$ is the deviation from the time mean at every grid point. Here, $f = [\overline{f}] + \overline{f}^* + f'$. In the present analysis, the time mean used to define the mean and stationary terms is taken over a 30-day window.

The storage term is written in terms of moist energy,

$$\text{ME} = C_v T + gz + Lq.$$

where $C_v$ is specific heat at constant volume. Following the standard energy budget (Held 2000; Donohoe et al. 2013), the prognostic storage involves ME, whereas the horizontal transport involves MSE.

Using these quantities, we first write the budget of zonally averaged, vertically integrated ME as

$$\frac{\partial}{\partial t}\text{VME}(y,t) + T_{\text{MSE}}(y,t) = F_{\text{MSE}}(y,t),$$

where

$$\text{VME}(y,t) = \int [\rho\,\text{ME}]\,\mathrm{d}z$$

is the density-weighted, vertically integrated moist energy, and

$$T_{\text{MSE}}(y,t) = \frac{\partial}{\partial y}\int [\rho v h]\,\mathrm{d}z$$

is the zonally averaged, vertically integrated meridional MSE transport term. MSE forcing is

$$F_{\text{MSE}}(y,t) = [Q_R + \text{SH} + \text{LH}],$$

where $Q_R$ is the column radiative heating, SH the surface sensible heat flux, and LH the surface latent heat flux.

The budget for the high-minus-low SST contrast of ME is then written as

$$\frac{\mathrm{d}}{\mathrm{d}t}\Delta\mathrm{VME}(t) + \Delta T_{\mathrm{MSE}}(t) + \mathrm{Residual} = \Delta F_{\mathrm{MSE}}(t),$$

where ΔVME, $\Delta T_{\mathrm{MSE}}$, and $\Delta F_{\mathrm{MSE}}$ are the differences between the subdomain means in the high-SST and low-SST regions. The residual term is computed from the other terms. In the sign convention used in the main text, a positive $\Delta F_{\mathrm{MSE}}$ mean $F_{\mathrm{MSE}}$ is larger in high-SST region than in low-SST region while a positive storage term means that ΔVME between the two regions increases with time. Because the domain is periodic, the meridional integral of $T_{\mathrm{MSE}}$ over the full domain is zero. The high-minus-low SST contrast $\Delta T_{\mathrm{MSE}}$ therefore measures the contrast between mean local export in the high-SST region and mean local import in the low-SST region, and is proportional to the net meridional transport from the high-SST region to the low-SST region. In the present simulations, the storage tendency is dominated by the latent-energy contribution $Lq$, while the contributions from $C_v T$ and $gz$ are typically at least one order of magnitude smaller.

In Fig. 8 of the main text, the mean and stationary eddy contributions are evaluated using 30-day analysis time windows for the early and late phases of the simulations (days 21–50 and 171–200, respectively). The residual term shown there represents the mismatch between the forcing, storage, and resolved transport terms. This residual is not negligible and likely reflects the combined effects of smaller-scale and/or more rapidly varying transport not fully captured by the present sampling (1 h), together with simplifications in the thermodynamic treatment such as the use of constant latent heat and specific heat for example. Nevertheless, the residual does not appear to change substantially across cases or time windows, so we do not expect it to affect the interpretation of the transport reorganization.